# Evolutionary Model of Stock Markets

Joachim Kaldasch

EBC Hochschule Berlin

Alexanderplatz 1, 10178 Berlin, Germany

(Email: joachim.kaldasch@international-business-school.de)

**Abstract**

The paper presents an evolutionary economic model for the price evolution of stocks. Treating a stock market as a self-organized system governed by a fast purchase process and slow variations of demand and supply the model suggests that the short term price distribution has the form a logistic (Laplace) distribution. The long term return can be described by Laplace-Gaussian mixture distributions. The long term mean price evolution is governed by a Walrus equation, which can be transformed into a replicator equation. This allows quantifying the evolutionary price competition between stocks. The theory suggests that stock prices scaled by the price over all stocks can be used to investigate long-term trends in a Fisher-Pry plot. The price competition that follows from the model is illustrated by examining the empirical long-term price trends of two stocks.

## 1. Introduction

Financial markets are one of the most intensive economic research fields. Great advantages have been achieved by analysing time series of the price and return of financial assets. Besides the standard asset price model based on geometric Brownian motion [1,2], stochastic volatility and multifractal models inspired by turbulence [3,4], multi-timescale, scaling and various types of self-similar theories [5-10] have been established. Also multi-agent models [11,12], Auto-Regressive-Conditional-Heteroskedastic (ARCH) and Generalized-ARCH (GARCH) models [13-15] are developed to describe the dynamics of financial markets [16,17].

The aim of this paper is to extend this mainstream research by establishing an evolutionary economic approach for the stock price evolution. The key idea of this model is that financial assets like other commodities are subject to competition in the trading process. This view has been emphasized in particular by Modis suggesting that stocks are in competition for investors' money [18]. He empirically showed that the stock price evolution can be considered to consist of a sequence of logistic growth periods [19]. Following Modis this finding indicates the price competition of stocks.

The presented consideration deviates from previous stochastic models by treating a stock market as a dynamic self-organized system in which the dynamics of demanded (desired) and supplied (available) units and the properties of the purchase process determine the price and return distributions of stock shares. The model is based on the idea that the purchase process of stock shares can be treated as a "reaction" between demanded and supplied units while the "reaction velocity" is the purchase rate. The purchase rate is in turn assumed to be governed by the law of mass action, i.e. it is proportional to the product of the number of available and demanded units. These ideas allow establishing price and return distributions of stocks which are in agreement with the main characteristics of empirical data. It also suggests that the long term mean price dynamics of stocks is governed by a Walrus equation [20]. Since this equation can be formulated as a replicator equation it can be shown that the long term price evolution of stocks suffers from competition. The focus of this paper is on a theoretical investigation of this price competition and a comparison with empirical stock price data.

The paper is organized as follows: In the second chapter the dynamic relations of a self-organized stock market are established. A short term price distribution that follows from the price dependent structure of demanded and supplied units and the corresponding purchase rate is derived. In the following chapters it is shown how the Walrus equation for the mean price of a stock can be obtained from perturbations of the distributions of demanded and supplied units and the transformation into a replicator equation. The next two chapters focus on the consequences of this relation. They discuss the price competition between stocks and show that the return distribution can be expected to have the form of Laplace-Gaussian mixture distributions. In order to compare the model with empirical data the price competition of two stocks is examined followed by a conclusion.

## 2. The Model

### 2.1. Stock Market Dynamics

The model is established for a closed speculative stock market. Agents, called investors (traders), purchase and sell stock units. In order to describe this market a number of variables have to be introduced. We want to characterize the number of investors by $A$ and indicate them with index $i$, while the number of stocks is termed $B$ and are indexed by $j$. The number of units (shares) of the $j$-th stock purchased per unit time by the $i$-th investor is indicated by $y_{ij}$. The number of units per unit time of the $j$-th stock sold by the $i$-th investor is denoted $y'_{ij}$. The total number of purchased units per unit time of the $j$-th stock $\tilde{y}_j$ must be equal to the total number of sold units per unit time $\tilde{y}_j'$ in the trading process[1]:

$$\tilde{y}_j = \sum_i^A y_{ij} = \sum_i^A y'_{ij} = \tilde{y}_j'$$
(1)

The purchase process is regarded as a number of stochastic events where demanded (desired) and supplied (available) shares meet and are exchanged between investors when they agree about the price $p$. The corresponding number of demanded shares of the $j$-th stock at a given price is denoted $n_j(p)$ and the number of supplied shares is $z_j(p)$.

We want to treat a stock market as a dynamic system that is governed by the dynamics of the number demanded and supplied units and the corresponding purchase rate. The dynamics can be given by conservation laws for both variables. The number of demanded units at a given price $p$ of the $j$-th stock is determined by:

$$\frac{dn_j(p)}{dt} = D_j(p) - y_j(p)$$
(2)

The number of demanded units increases with the demand rate $D_j(p)$ and decreases with the purchase rate $y_j(p)$. The demand rate $D_j(p)$ characterizes the generation rate of desired units per unit time by investors at a given price. It can be regarded to be a function of many variables. Among them is for example the information flow, risk adversion of investors, the expected profitability of invested money etc.

Also the number of available units $z_j(p)$ offered for a price $p$ is governed by a conservation relation:

$$\frac{dz_j(p)}{dt} = S_j(p) - y_j(p)$$
(3)

The number of available units increases with the supply rate $S_j(p)$ and decreases when units are purchased with the purchase rate $y_j(p)$. The supply rate $S_j(p)$ represents the generation rate of available shares per unit time.

Further introduced are the total numbers of demanded and supplied units by:

---

[1] Variables and functions related to numbers of stock shares are regarded to be scaled by a large number such that they can be treated as real numbers.

$$\tilde{n}_j = \int\limits_0^\infty n_j(p)dp; \tilde{z}_j = \int\limits_0^\infty z_j(p)dp$$

(4)

while the total purchase rate of the *j-th* stock given by Eq.(1) can be obtained from:

$$\tilde{y}_j = \int\limits_0^\infty y_j(p)dp$$

(5)

The key idea of the presented model is to regard the purchase process as governed by the law of mass action. That means, we assume that the purchase rate $y_j(p)$ must be proportional to the product of the number of supplied units $z_j(p)$ and the number of demanded units $n_j(p)$.[2] In other words the purchase process is interpreted as a "reaction" between demanded and supplied units where the "reaction velocity" corresponds to the purchase rate of the form:

$$y_j(p) \cong \eta_j n_j(p) z_j(p)$$

(6)

The rate $\eta_j$ is termed preference rate. It characterizes the chance that just the *j-th* stock is chosen by investors in a given time period.[3]

In order to describe the self-organized dynamics of the stock market we take advantage of the synergetic approach and separate the processes with respect to their velocity [21,22]. In the self-organized view of a dynamic system fast processes relax rapidly to their stationary states such that over long time periods the fast process is "slaved" by slow processes. For a stock market the purchase process is regarded as the fast process while variations of the demand and supply rates are treated as the slow process. In order to take this effect into account we introduce a separation of the time scales. The slow time scale is denoted by $t$ while $\tau$ indicates the fast time scale, while $dt >> d\tau$. On the short time scale the demand and supply rates can be treated as constant:

$$\frac{dD_j(p)}{d\tau} = \frac{dS_j(p)}{d\tau} \cong 0$$

(7)

Due to the fast purchase process the distributions of demanded and supplied units relax rapidly towards their stationary states. They can be obtained from Eqs.(2)-(3) by setting:

$$\frac{dn_j(p)}{d\tau} = \frac{dz_j(p)}{d\tau} = 0$$

(8)

This leads to:

$$D_j(p) = S_j(p) = y_j(p)$$

(9)

---

[2] This is a first order approximation for $y(p)$, since the purchase rate must be zero if either $n(p)$ or $z(p)$ disappear.
[3] Here the rate $\eta$ plays the role of an "affinity", which is in economic terms a preference characterizing the mean trading activity of the stock.

As shown in Appendix *A*, the stationary solution Eq.(9) corresponds to a stable state if the demand and supply rates can be treated as constant. In the stationary state the purchase rate (unit sales) and the supply and demand rates are equal.

## 2.2. The Price Distribution

The model suggests that the purchase process determines the stock market dynamics on the short time scale. However, the purchase process is governed by its own laws. The aim of this chapter is to establish the stationary probability density function (pdf) $P_y(p)$ characterizing the relative abundance that purchase events of a stock occur within a price interval $p$ and $p+dp$. This distribution is termed price distribution and can be obtained from[4]:

$$P_y(p) = \frac{1}{\bar{y}} y(p)$$

(10)

with the cumulative distribution function (cdf):

$$F_y(p) = \int_0^p P_y(p') dp'$$

(11)

Because the price is strictly positive we further demand:

$$F_y(p) = \begin{cases} 0 & for \quad p = 0 \\ 1 & for \quad p \to \infty \end{cases}$$

(12)

The mean price $\mu$ that results from the purchase process is determined by:

$$\mu = \int_0^\infty P_y(p) p \, dp$$

(13)

The probability density functions of the stationary number of demanded and supplied units are given by:

$$P_n(p) = \frac{1}{\tilde{n}} n(p)$$

(14)

and

---

[4] Focussing on a single stock the index is omitted.

$$P_z(p) = \frac{1}{\tilde{z}} z(p)$$

(15)

while the cumulative distribution functions (cdf) become:

$$F_n(p) = \int_0^p P_n(p')dp'; F_z(p) = \int_0^p P_z(p')dp'$$

(16)

For the derivation of the price distribution $P_y(p)$ we demand that sufficiently many purchase events occur within a short time period such that the probability functions introduced above are in their stationary state. Since all units of a stock are equivalent, investors prefer to purchase stocks for the lowest available price. Hence, the number of demanded units $n(p)$ increases for $p \rightarrow 0$ and must have its maximum $\tilde{n}$ at the lowest price at which purchase events occur. The number of demanded units at $p$ is therefore determined by:

$$n(p) = \tilde{n} - \tilde{n} \int_0^p P_n(p')dp' = \tilde{n}(1 - F_n(p))$$

(17)

which decreases with an increasing price. On the other hand the number of available units $z(p)$ will increase with an increasing price and approaches its maximum $\tilde{z}$ for $p \rightarrow \infty$. The number of supplied units for investors who wants to pay a price $p$ is determined by all units up to this price. Hence, the number of available units $z(p)$ is given by:

$$z(p) = \tilde{z} \int_0^p P_z(p')dp' = \tilde{z} F_z(p)$$

(18)

which increases monotonically with increasing price. For the purchase rate Eq.(6) we obtain with Eq.(17) and Eq.(18):

$$y(p) = \eta \tilde{z} F_z(p) \tilde{n}(1 - F_n(p))$$

(19)

With Eq.(5) the total sales become:

$$\tilde{y} = \eta \tilde{z} \tilde{n} \int_0^\infty F_z(p)(1 - F_n(p))dp = \eta \tilde{z} \tilde{n} T$$

(20)

The price distribution Eq.(10), along with Eq.(19) and Eq.(20), therefore has the form:

$$P_y(p) = \frac{1}{T} F_z(p)(1 - F_n(p))$$

(21)

A monotone decreasing and a monotone increasing function have a single interception point $p*$ at which:

$$n(p*) = z(p*)$$
(22)

For a price $p'' < p*$, the number of desired units increases the number of available units: $n(p'') > z(p'')$. Therefore the purchase rate is limited by the number of available units $z(p'')$. The relative abundance of purchase events in the price interval $p''$ and $p'' + dp$ is therefore approximately equal to the relative abundance of available units: $P_y(p'') \approx P_z(p'')$. For very small prices the chance to find demanded units is: $1 - F_n(p'') \approx 1$. Hence, the price distribution Eq.(21) can be approximated for $p'' << p*$ by:

$$P_y(p'') \cong \frac{1}{T} F_z(p'') = \frac{1}{T} \int_0^{p''} P_z(p') dp' \cong \frac{1}{T} \int_0^{p''} P_y(p') dp' = \frac{1}{T} F_y(p'')$$
(23)

On the other hand, for $p'' > p*$, it can be argued that the number of available units increases the number of desired units: $n(p'') < z(p'')$. Purchase events are therefore limited by the number of desired units and the relative abundance of purchase events are approximately equal in this case to the relative abundance of desired units: $P_y(p'') \approx P_n(p'')$. For very large prices we can further approximate the chance to find supplied units by $F_z(p) \approx 1$ and Eq.(21) becomes, for $p'' >> p*$:

$$P_y(p'') \cong \frac{1}{T} \left( 1 - F_n(p'') \right) = \frac{1}{T} \left( 1 - \int_0^{p''} P_n(p') dp' \right) \cong \frac{1}{T} \left( 1 - \int_0^{p''} P_y(p') dp' \right) = \frac{1}{T} \left( 1 - F_y(p'') \right)$$
(24)

In order to satisfy the boundary relations Eq.(23) and Eq.(24), the stationary price distribution can be estimated by:

$$P_y(p) \cong \frac{1}{T} F_y^s(p) \left( 1 - F_y^s(p) \right) = P_y^s(p)$$
(25)

with the approximated price distribution $P_y^s(p)$ and the corresponding cdf $F_y^s(p)$. Note that $P_y^s(p)$ is a symmetric distribution around the mean price $\mu$. Since the distributions of demanded and supplied units Eq.(17) and Eq.(18) turn with this approximation into:

$$z(p) \cong \tilde{z} F_y^s(p); n(p) \cong \tilde{n} \left( 1 - F_y^s(p) \right)$$
(26)

we obtain for the scaling variable $T$ :

$$T = \int_0^\infty F_z(p) \left( 1 - F_n(p) \right) dp \cong \int_0^\infty F_y^s(p) \left( 1 - F_y^s(p) \right) dp = \int_0^\infty P_y^s(p) dp = 1$$
(27)

while we used the definition Eq.(12). Realizing that:

$$P_y^{\;s}(p) = \frac{dF_y^{\;s}(p)}{dp} = F_y^{\;s}(p)\left(1 - F_y^{\;s}(p)\right)$$

(28)

is a logistic differential equation the cdf of the price distribution must have the form:

$$F_y^{\;s}(p) = \frac{1}{1 + e^{-\frac{p-\mu}{Q'}}}$$

(29)

where $Q'$ is a free parameter. In order to determine this parameter we assume that there is a lowest price $\mu_m$ at which purchase events occur. At this price is [5]:

$$F_y^{\;s}(\mu_m) = \varepsilon'$$

(30)

with $\varepsilon' << 1$. The parameter $Q'$ in Eq.(29) becomes with Eq.(30):

$$Q' = Q(\mu - \mu_m)$$

(31)

while:

$$Q = \frac{1}{\ln\left(\frac{1}{\varepsilon'} - 1\right)}$$

(32)

The symmetric price distribution can be given by a logistic distribution of the form:

$$P_y^{\;s}(p) = \frac{1}{Q(\mu - \mu_m)} \frac{e^{\frac{p-\mu}{Q(\mu - \mu_m)}}}{\left(1 + e^{\frac{p-\mu}{Q(\mu - \mu_m)}}\right)^2}$$

(33)

with the variance:

$$Var(P_y^{\;s}) = \frac{Q^2(\mu - \mu_m)^2 \pi^2}{3}$$

(34)

---

[5] It is assumed that below the cut-off price $\mu_m$ purchase events disappear because they are not profitable for the suppliers and $\varepsilon'$ is sufficiently small to guarantee Eq.(12).

Confining the presented model to stocks with mean prices $\mu >> \mu_m$, the price distribution can be further simplified. Reducing Eq.(33) to the exponential tails the price distribution is a Laplace distribution of the form:

$$P_y^s(p) \cong \frac{1}{2q\mu} e^{-\frac{|p-\mu|}{q\mu}}$$

(35)

with[6]:

$$q = \frac{Q\pi}{\sqrt{6}}$$

(36)

The normalization constant $q$ can be related to the empirical standard deviation $\sigma$ of price fluctuations by:

$$\sigma = \sqrt{2}q\mu$$

(37)

Schematically displayed in Fig.1 are the functions $n(p)$ and $z(p)$ as suggested above (dashed lines). Investors willing to pay a price $p''$ have the choice between $z(p'')$ available units and represent $n(p'')$ desired units. The discrepancy between them is indicated by the arrow in Fig.1. The purchase rate is limited to $z(p'')$ for $p''<<\mu$ and to $n(p'')$ for $p''>>\mu$. Only at mean price $\mu$ both are equal in this symmetric constellation. The logistic price distribution $P_y^s(p)$ is displayed by the solid line in Fig.1 and the dotted line indicates the corresponding Laplace distribution.

Note that the standard deviation of the price distribution is in this model not a consequence of the central limit theorem. The width of the price distribution can rather be interpreted as a degree of consensus about bid and ask prices of demanded and supplied stock units. If sellers and buyers mainly agree about the price of a stock the standard deviation will be small and hence we get a sharp price distribution. If they largely disagree, the price of offered and sold units will spread leading to a more dispersed purchase price distribution.

## 2.3. The Mean Price Evolution

Demand and supply rate fluctuations may disturb the price distribution Eq.(33) such that the resulting price distribution is asymmetric. We can therefore generally write the total price distribution as a symmetric contribution $P_y^s(p)$ and a perturbation $\delta P(p)$:

$$P_y(p) = P_y^s(p) + \delta P(p)$$

(38)

In order to keep the model simple we want to confine the following considerations to the case that perturbations of the symmetric price distribution can be regarded as small:

---

[6] This relation between $Q$ and $q$ is based on the idea that the price standard deviations of both distributions are equivalent.

$$\delta P(p) << P_y^s(p)$$
(39)

In this case the stationary price distribution can be approximated by the symmetric distribution as displayed schematically in Fig.1:

$$P_y(p) \cong P_y^s(p)$$
(40)

The price distribution can then be described by two free parameters: the price variance and the mean price. The approximation Eq.(40) further demands that the interception point of demanded and supplied units is nearly equal to the mean price: $p^*\approx\mu$. Since $F(\mu)=1/2$, we obtain from Eq.(22) with Eq.(26) that:

$$\tilde{z} \cong \tilde{n}$$
(41)

In this chapter we want to consider the price evolution on the long time scale. On this time scale the demand and supply rates are not constant as suggested by Eq.(7). As a consequence also the number of demanded and supplied units $n(p)$ and $z(p)$ change in time. But because the purchase process takes place much faster than variations of the demand and supply rates, the perturbed functions of demanded and supplied units $n'(p)$ and $z'(p)$ re-establish are after a short relaxation period such that price distribution becomes stationary as given by Eq.(33). The perturbation leads, though, to a new mean price $\mu'=\mu+\delta\mu$. The new intersection point of the new demanded and supplied units $n'(\mu')$ and $z'(\mu')$ has to satisfy Eq.(22). Hence:

$$n'(\mu') = z'(\mu')$$
(42)

The new distributions of demanded and supplied units can be written as perturbations of the old distributions:

$$n'(\mu') = n(\mu') + \delta n(\mu); z'(\mu') = z(\mu') + \delta z(\mu)$$
(43)

with the perturbations $\delta n(\mu)$ and $\delta z(\mu)$. Expanding the functions:

$$n(\mu') = n(\mu) + \left.\frac{dn(p)}{dp}\right|_\mu \delta\mu; z(\mu') = z(\mu) + \left.\frac{dz(p)}{dp}\right|_\mu \delta\mu$$
(44)

we get with Eq.(26):

$$\left.\frac{dn(p)}{dp}\right|_\mu = \left.\frac{d\tilde{n}(1-F_y(p))}{dp}\right|_\mu = -\tilde{n}\left.\frac{dF_y(p)}{dp}\right|_\mu; \left.\frac{dz(p)}{dp}\right|_\mu = \tilde{z}\left.\frac{dF_y(p)}{dp}\right|_\mu$$
(45)

while Eq.(33) suggests that:

$$\left.\frac{dF_y(p)}{dp}\right|_\mu = P_y(\mu) = \frac{1}{4Q\mu}$$
(46)

Applying the relations Eq.(41-46) we obtain for the price change:

$$\frac{\delta\mu}{\mu} = \frac{2Q}{\tilde{n}}\big(\delta n(\mu) - \delta z(\mu)\big)$$
(47)

For a change of the mean price during on a long time interval $dt$ the relation turns into:

$$\frac{1}{\mu}\frac{d\mu}{dt} = H\left(\frac{dn(\mu)}{dt} - \frac{dz(\mu)}{dt}\right) = H\big(D(\mu) - S(\mu)\big)$$
(48)

where we used Eq.(2) and Eq.(3) and:

$$H = \frac{2Q}{\tilde{n}}$$
(49)

Eq.(48) is known as the Walrus equation [20]. It suggests that an excess demand rate $D(\mu)$ increases the mean price, while an excess supply rate $S(\mu)$ decreases the mean price during long time intervals[7].

## 2.4. The Replicator Dynamics of Stock Prices

While Eq.(9) suggests that the mean price is constant for short time periods, Eq.(48) implies that the mean price of the *j-th* stock evolves on the long time scale as:

$$\frac{1}{\mu_j(t)}\frac{d\mu_j(t)}{dt} = f_j(t)$$
(50)

where we introduced the growth rate:

$$f_j(t) = \frac{Q_j}{n_j(t)}\big(D_j(t) - S_j(t)\big)$$
(51)

It is governed by the evolution of the supply and demand rates.

Further we want to introduce the sum over all mean stock prices in the market and denote it as the total price:

---

[7] In the neo-classic theory the equivalence of demand and supply flows expresses market equilibrium. The mean price $\mu$ is in this view the so-called equilibrium price.

$$\tilde{\mu}(t) = \sum_{j=1}^{B} \mu_j(t)$$

(52)

From Eq.(50) follows that the total price evolves as [8]:

$$\frac{d\tilde{\mu}}{dt} = \sum_j f_j \mu_j = \langle f \rangle_\mu \tilde{\mu}$$

(53)

Further introduced is the relative stock price $w_j(t)$ by:

$$w_j(t) = \frac{\mu_j(t)}{\tilde{\mu}(t)}$$

(54)

Taking the time derivative from Eq.(54) we obtain:

$$\frac{dw_j}{dt} = \frac{1}{\tilde{\mu}} \frac{d\mu_j}{dt} - \frac{\mu_j}{\tilde{\mu}^2} \frac{d\tilde{\mu}}{dt} = \left( f_j - \langle f \rangle_\mu \right) w_j = \Delta f_j w_j$$

(55)

where we used Eq.(50) and Eq.(53). The Walrus equation turns into a replicator equation for relative stock prices. The replicator relation is known from evolutionary theories to describe competition while the function $f$ is called fitness [22]. In this case the presented model suggests that relative stock prices are in an evolutionary competition governed by the fitness function $f$ which we want to denote here as stock fitness, and $\Delta f_i$ is the corresponding fitness advantage of a stock. Since the fitness function is determined by Eq.(51) we can conclude that the stock price fitness is essentially governed by the evolution of demand and supply rates.

The stationary solution of Eq.(55) is:

$$w_j = \begin{cases} 1 & if \; \Delta f_j = 0 \\ 0 & if \; \Delta f_j \neq 0 \end{cases}$$

(56)

This result suggests that in the case of large growth rates $\Delta f_j$ stocks would replace each other very fast with respect to the relative price until just one stock survives with $f_j = <f>_\mu$ and $w_j = 1$. Since this is usually not the case in stock markets it can be concluded that the price growth rate of stocks must be small, i.e. $\Delta f_j << 1$.

The time evolution of the relative price can be obtained by writing the replicator equation as:

$$\ln(w_j) = \Delta f_j dt$$

(57)

Diminishing the relative price by a second stock with index $l$ we obtain:

---

[8] Terms in brackets are functions averaged with respect to the variable in the index.

$$\ln\left(\frac{w_j}{w_l}\right) = \Delta f_{jl}\Delta t + C$$

(58)

where $C$ is a constant of integration and the fitness advantage is a time average over the fitness difference:

$$\Delta f_{jl} = \left\langle f_j(t) - f_l(t)\right\rangle_{\Delta t} = \int_{\Delta t}\left(f_j(t) - f_l(t)\right)dt$$

(59)

The time evolution of the relative price becomes:

$$w_j(t) \sim w_l(t)e^{\Delta f_{jl}\Delta t}$$

(60)

Displaying the relative price in a half-logarithmic plot as a function of the time $\Delta t$ (Fisher-Pry plot), linear relationships in the price evolution are associated with a constant competitive (dis-)advantage $\Delta f_{jl}$. They have their origin according to Eq.(51) in a permanent excess demand (supply) rate for shares of the $j$-$th$ stock.

We can also calculate the fitness advantage against all other stocks. In this case is:

$$w_l(t) = 1 - w_j(t)$$

(61)

and Eq.(60) can be approximated for small $w_j$ by:

$$\mu_j(t) \sim \tilde{\mu}(t)e^{\Delta f_j\Delta t}$$

(62)

If the total price is nearly constant the relation suggests the appearance of an exponential growth in the price evolution for $\Delta f_j > 0$ as a result of the competitive advantage between the stock and the market. A consequence of the price competition is that a permanent disadvantage may lead to the disappearance of a stock. Over a sufficiently long time period the preferential growth process therefore causes an adaptation of the variety of stocks in the market.

## 2.5. The Return Distribution

In this section we want to discuss the long-term return distribution that follows from the presented model of the stock price evolution. Price variations occurring at different time steps $t$ and $t+\delta t$ are related to the return $r$ of a stock by[9]:

---

[9]Confining the study to a single stock, the index is omitted in the following considerations.

$$r = \frac{p(t + \delta t) - p(t)}{p(t)} = \frac{p(t + \delta t)}{p(t)} - 1$$

(63)

For sufficiently small time steps $\delta t$ the return is usually a small quantity and we can approximate:

$$r \cong \ln(r + 1) = \ln\left(\frac{p(t + \delta t)}{p(t)}\right)$$

(64)

The mean stock price evolution can be considered to consist of a time averaged and a fluctuating growth rate contribution:

$$f(t) = \langle f \rangle_t + \delta f(t)$$

(65)

Under the condition that fitness fluctuations induce uncorrelated mean price fluctuations:

$$\langle \delta\mu(t), \delta\mu(t') \rangle_t \sim \delta(t - t')$$

(66)

the mean price evolution Eq.(50) can be formulated as a stochastic differential equation of the form:

$$d\mu \cong \langle f \rangle_t \mu dt + \sigma' \mu dW$$

(67)

where $\sigma'$ is the standard deviation of mean price variations and the differential $dW$ indicates a iid random variable with null mean and variance equal to $dt^{1/2}$. The central limit theorem suggests that the mean price evolution is governed in this case by a multiplicative random growth process. It leads after sufficient time to a lognormal distribution of the mean price:

$$P_\mu(\mu) = \frac{1}{\sqrt{2\pi}\omega\mu} \exp\left(-\frac{\left(\ln(\mu/\mu^0) - \rho\right)^2}{2\omega^2}\right)$$

(68)

where $\rho$ and $\omega$ are free parameters and $\mu/\mu^0$ is the mean stock price at $t$ scaled by the price at a time step $t_0$.

The model suggests however that the price of a stock on the long time scale is determined by of two processes one related to the fast purchase process generating the symmetric price distribution Eq.(33) and the other by much slower demand and supply rate variations generating the mean price distribution Eq.(68). In order to take both effects into account an unconditional price distribution $P_y(p)$ is established by:

$$P_y(p) = \int_0^\infty P_y(p \mid \mu) P_\mu(\mu) d\mu$$

(69)

where the conditional price distribution $P_y(p|\mu)$ is characterizing the probability density of a stock price $p$ under the condition that the mean stock price is just $\mu$.

We want to confine the discussion to two extreme cases:

1. In the first case the purchase price distribution Eq.(33) is regarded to be a very narrow distribution located around the mean price[10]. The conditional price distribution $P_y(p|\mu)$ can then be approximated by a Dirac-delta function of the form:

$$P_y(p \mid \mu) \approx \delta(p - \mu)$$

(70)

and the unconditional price distribution becomes:

$$P_y(p) = \int_0^\infty \delta(p - \mu) \frac{1}{\sqrt{2\pi}\omega\mu} \exp\left(-\frac{\left(\ln(\mu / \mu^0) - \rho\right)^2}{2\omega^2}\right) d\mu = \frac{1}{\sqrt{2\pi}\omega p} \exp\left(-\frac{\left(\ln(p / p^0) - \rho\right)^2}{2\omega^2}\right)$$

(71)

That means, even if variations of the demand and supply rates are slow, they are the only relevant price fluctuations and therefore govern the price distribution. Writing $p/p^0 = p(t_0 + \delta t)/p(t_0) = r+1$ we can take advantage from Eq.(64) and obtain for the return probability density distribution by a change of the variables in Eq.(71) a normal return distribution of the form:

$$P_N(r) = \frac{1}{\sqrt{2\pi}\omega} \exp\left(-\frac{(r - \rho)^2}{2\omega^2}\right)$$

(72)

where $\rho$ and $\omega$ are the mean return and return standard deviation, respectively. The geometric Brownian motion of the mean price generates a normal return distribution. This case corresponds to the standard model of price returns as suggested by Osborn [23].

2. In the second case we assume that demand and supply rate variations are either very slow or can be regarded to disappearing. The mean price is in this case a constant $\mu_0$ and the distribution Eq.(68) is located around this constant such that:

$$P_\mu(\mu) \approx \delta(\mu - \mu_0)$$

(73)

Taking advantage from the approximation Eq.(35) for the tails of the price distribution we can further write for the unconditional price distribution Eq.(69):

$$P_y(p) = \int_0^\infty \frac{1}{2q\mu} e^{-\frac{|p-\mu|}{q\mu}} \delta(\mu - \mu_0) d\mu \cong \frac{1}{2q\mu_0} e^{-\frac{1}{q}\left|\frac{p-\mu_0}{\mu_0}\right|}$$

(74)

---

[10] This can be expected for penny stocks.

Hence the price distribution is completely determined by the fast purchase process. Introducing the scaled price variable:

$$\kappa + 1 = \frac{p}{\mu_0}$$

(75)

the Laplace distribution Eq.(74) turns by a change of the variables into:

$$P(\kappa) = \frac{1}{2q} e^{-\frac{1}{q}|\kappa|}$$

(76)

Under the condition that $\delta p = p(t) - \mu_0 << 1$ the function $\kappa = \delta p / \mu_0$ is a small quantity. Taking the logarithm of Eq.(75) we can approximate:

$$\ln\left(\frac{p}{\mu_0}\right) = \ln(\kappa + 1) \cong \kappa$$

(77)

For two time steps $t_2, t_1$ with $t_2 > t_1$, the difference between two scaled random prices $\kappa(t_2)$ and $\kappa(t_1)$ at constant mean price is then related to a return:

$$r(t_2 - t_1) = \kappa(t_2) - \kappa(t_1) = \ln\left(\frac{p(t_2)}{\mu_0}\right) - \ln\left(\frac{p(t_1)}{\mu_0}\right) = \ln\left(\frac{p(t_2)}{p(t_1)}\right)$$

(78)

Under the condition that prices at two different time steps can be treated as independent the return distribution is the difference of the price distribution Eq.(76) at two different time steps:

$$P_L(r) = \int_{-\infty}^{\infty} P(\kappa) P(\kappa - r) d\kappa = \frac{1}{4q^2} \int_{-\infty}^{\infty} e^{-\frac{1}{q}|\kappa|} e^{-\frac{1}{q}|\kappa - r|} d\kappa \cong \frac{1}{2q} e^{-\frac{1}{q}|r|}$$

(79)

where we used that $r$ is a small quantity. For neglecting demand and supply rate fluctuations the return distribution has therefore the form of a Laplace distribution while $q$ can be specified by the return standard deviation:

$$\sigma_r = \sqrt{2} q$$

(80)

Note that the return distributions Eq.(76) and Eq.(79) express extreme cases, in the sense that the impact of one process characterizing the unconditional price distribution is neglected. In general the return distribution is a mixture of $P_N(r)$ and $P_L(r)$. Several variants of Laplace-Gaussian mixtures have been proposed in the literature [24]. A well-known distribution that nests the Laplace and the normal distribution is the generalized exponential distribution (GED) [25]. With standardized density and zero-mean return it has the form:

$$P(r,\lambda) = \frac{2^{-(1/\lambda+1)}\lambda}{\Gamma(\lambda^{-1})} e^{-\frac{1}{2}|r|^\lambda}$$

(81)

where $\lambda > 0$ is a shape parameter and $\Gamma$ indicates the Gamma function. For $\lambda = 1$, the distribution reduces to the Laplace distribution and for $\lambda = 2$ the normal distribution is obtained. An even more general distribution is the hyperbolic distribution. It can be written with zero location and standard variance as [26]:

$$P(r,\lambda,\chi) = C_1 e^{C_2\left(\lambda\sqrt{1+r^2} - \chi r\right)}$$

(82)

while

$$C_1 = \frac{\sqrt{\lambda^2 - \chi^2}}{2\lambda K_1(\lambda^{-2}-1)}; C_2 = \frac{\lambda^2 - 1}{\lambda^2\sqrt{\lambda^2 - \chi^2}}$$

(83)

and $K_1$ is the modified Bessel function of the third kind with index one. The hyperbolic distribution also nests the Gaussian and the Laplace distribution. For $\lambda = 0$ the density function is symmetric. For $\chi \to 0$ the return distribution approaches a normal density and for $\lambda = 0$, $\chi = 1$ it coincides with a scaled Laplace distribution.

Another combination is a simple mixture of Gauss and Laplace densities as was suggested by Kanji [27]:

$$P(r,\theta) = \theta P_N(r) + (1-\theta)P_L(r)$$

(84)

In this approach the mixing proportion $0 \leq \theta \leq 1$ determines the impact of the Gaussian respectively Laplace distribution on the return.

Also suggested as a Laplace-Gaussian mixture is a Gauss-Laplace sum as the weighted sum of a random normal variable $N'$ and a Laplace distributed variable $L'$ [28]:

$$Z = \theta N' + (1-\theta)L'$$

(85)

All these models are based on a constant variance. Except from the cases Eq.(72) and Eq.(79) it has to be emphasized, however, that the standard deviation of the price distribution is a function of the mean price as suggested by Eq.(34). Since the mean price $\mu(t)$ is a fluctuating variable the price variance must be also a fluctuating variable. This effect is known from time series analysis and can be taken into account by applying ARCH dynamics in the price evolution.

It is further known that human activities are not uncorrelated. Empirical investigations suggest that they occur in bursts and the time evolution of human activities is governed by long term correlations [29]. As a consequence the mean price evolution which is the result of supply and demand rate variations in this model cannot be treated as uncorrelated as suggested by Eq.(66). Correlated price variations instead induce a pronounced mountain-valley structure of the price, where statistically large values are likely to be followed by large values and small by small ones.[11] Even jumps of the mean price may occur (bubbles/crashes). These effects are termed volatility clustering and can be modelled in the price analysis of financial assets for example by Generalized-ARCH (GARCH) type price dynamics. Correlated price variations therefore cause deviations from the Laplace-Gaussian mixture distributions. They can be taken into account for example by including GARCH-type dynamics with mixtures of Laplace-Gaussian random variables [28].

### 3. Comparison with Empirical Results

The presented economic model makes two main statements:

1.) The symmetric part of the price distribution of a stock can be described for a constant mean price by a logistic distribution which can be approximated by a Laplace distribution. Taking small mean price fluctuations into account the return distribution turns into a Laplace-Gaussian mixture distribution.

2.) The model suggests that the stock price evolution is governed by the mutual competition between stocks. A constant competitive advantage leads to long term linear trends of the relative stock price in a Fisher –Pry plot.

The first statement implies that return distributions have generally the structure of Laplace-Gaussian mixture distributions. The Laplace distribution was already applied for example by Kotz et al. to describe the empirical return distributions [30]. An extensive empirical analysis applying Laplace-Gaussian mixture distributions including GARCH-type dynamics has been performed by Haas et al.[28]. They found a good agreement of this type of distributions with empirical data and reported that applying Laplace-Gaussian mixture distributions with GARCH-type dynamics is competitive with and even superior to the use of hyperbolic distributions.

Here we want to confine to two stocks traded at the NYSE: Apple and Microsoft. They are compared to the Dow Jones Industrial Average (DJIA) which is treated here as a measure of the total price of the stock market (The corresponding data are available at Yahoo.com Finance). Displayed in Fig.2 is the long term evolution of the relative price of the two stocks in relation to the DJIA. Their return distributions are shown in the insert. As an illustration the central part of the return distributions are fitted by a Laplace-Gaussian mixture distribution. The solid lines indicate a GED as given by Eq.(81) with $\lambda_{Apple} = 1.2$, $\sigma_{Apple} = 0.015$, $\lambda_{Microsoft} = 1.1$, $\sigma_{Microsoft} = 0.001$. Both distributions are nearly Laplace distributions in agreement with previous empirical results [31]. [12]

The model suggests that setting out the mean price as a function of time linear relationships in the Fisher-Pry plot reveal competitive (dis-)advantages in the evolutionary price competition between the two stocks and the DJIA[13]. And indeed the Microsoft stock shows an almost linear trend in the Fisher-Pry plot from the beginning until 1999 indicated by the regression line. It suggests that Microsoft shares have in this period a nearly constant evolutionary advantage against the DJIA. However, in 1999 Microsoft enters the DJIA and contributes to its price with a fixed share. Hence the competition disappears on average as indicated by the constant regression line.

---

[11] These price variations are also known in speculative markets as Elliott waves.

[12] The model is applicable if the price and hence the return distributions are symmetric and can be described by a Laplace-Gaussian mixture distribution. The empirical skewness is: $s_{Apple} = -0.1$, $s_{Apple} = -0.3$. Excluded from the presented return distribution are the transition periods between the linear trends in the Fisher-Pry plot.

[13] Note that the model is based on the mean price of a stock while the empirical price data represent the last purchase event of a day. The empirical data are taken as a proxy for the mean price.

This never happened with Apple shares. The price evolution exhibits periods of competitive advantages and disadvantages. In particular in the period 1988-1998 a negative trend in the price evolution is evident shown by the first regression line in Fig. 2. The trend broke around 1999 is accompanied with the re-entering of the firm founder Steve Jobs. The competitive advantage against the DJIA (and hence the mean stock market) in the following years is due to a permanent excess demand rate. The excess demand can be interpreted as a result of affirmed collective expectations of investors about the firm growth. Note that linear trends in the Fisher-Pry-plot correspond to logistic growth periods in a price chart. It is therefore also possible to interpret successive linear price trends in the Fisher-Pry-plot as a sequence of long term logistic growth periods in the price chart of a stock if the impact of the total price evolution can be neglected. This was suggested already by Modis [18,19][14].

## 4. Conclusion

The dynamic evolutionary theory established above suggests that a stock market can be viewed as a self-organized system in which the price dynamics of stocks is determined on the one hand by a fast purchase process and on the other hand by slow variations of the demand and supply rates (i.e. of the generation rate of new demanded and supplied stock shares). The distributions of demanded units $n(p)$ and supplied units $z(p)$ have in this model the form schematically displayed in Fig.1. The graph is reminiscent of pictures of the neo-classic economic theory where the intersection point of aggregated demand and supply curves determine market equilibrium [17]. And indeed since the distributions of supplied and demanded units $z(p)$ and $n(p)$ are determined by cumulative price distributions, they can be interpreted in terms of the neo-classic approach as aggregated demand and supply curves of a stock. But in difference to the neo-classic theory these curves are not established from utility considerations. They are the result of the purchase process.

The key idea of the presented theory is that the price dependent purchase rate describing the purchase process is proportional to the product of number demanded and supplied units: $y(p) \sim n(p)z(p)$. The intersection point of $n(p)$ and $z(p)$ expressing maximum coincidence between desired and available stock shares must therefore be the maximum of the purchase rate $y(p)$ and thus of the price distribution $P_y(p)$ also displayed in Fig.1. The theory is confined to a study of symmetric price distributions where the mean price is equivalent to the intersection point of demanded and supplied units. In economic terms the mean price corresponds then to the so-called equilibrium price.

Slow variations of the demand and supply rates $D(p)$ and $S(p)$ changes the aggregated demand and supply curves and therefore the price distribution. But the purchase process is treated in this model as a much faster and therefore the symmetric price distribution is re-established after a short relaxation but at a different mean price (and varying variance). The price of a share can therefore be regarded to be governed by two stochastic processes. On is related to the fast purchase process and the other to slow stochastic variations of the demand and supply rates. If the impact of the latter can be neglected it is shown that the price distribution has the form of a logistic distribution as displayed in Fig.1 while the return distribution can be approximated by a Laplace distribution. If, however, the price distribution is localized and uncorrelated demand and supply rate fluctuations dominate, the price distribution becomes lognormal and the return distribution turns into a Gaussian distribution. In general the return distribution can therefore be expected to have the form of Laplace-Gaussian mixture distributions. This was found in empirical investigations [28]. Human activities are in difference to the model assumptions correlated. Therefore the price dynamics has to be generalized by GARCH-type price dynamics not performed here.

---

[14] See also additional examples in this reference.

Explicitly derived in this theory is the Walrus equation for evolution of the mean price. This relation implies that the mean price of a stock increases with an excess demand rate and decreases with excess supply rate of a stock. The model suggests that the evolution of the mean price of a stock scaled by the sum over all stock prices can be described by a replicator equation. The replicator dynamics generates a price competition between stocks that can be studied in a Fisher-Pry–plot. Linear trends in this plot indicate a permanent competitive advantage compared to other stocks.

Discussed as an illustration are the long term price dynamics of two stocks: Microsoft and Apple. It is shown that similar to commodities or technologies a presentation of the price evolution of stocks in a Fisher-Pry plot is a useful tool to study long term trends induced by the mutual price competition between stocks and between a stock and the total market respectively.

**Appendix A**

The stationary solution of Eq.(2) and Eq.(3) is given by Eq.(9). For an examination of the stability of this stationary state we write the number of demanded and supplied units as:

$$n(\tau) = n_0 + \delta n(\tau)$$
$$z(\tau) = z_0 + \delta z(\tau)$$
(A1)

where $n_0$ and $z_0$ are the stationary states. Since the demand and supply flows vary slowly on this time scale we can write:

$$D = S = \eta z_0 n_0$$
(A2)

Inserting these relations in Eq.(2) and Eq.(3) we obtain for the time dependent fluctuations around the stationary state:

$$\frac{d\delta n}{d\tau} = -\eta \left( n_0 \delta z + z_0 \delta n + \delta n \delta z \right)$$
$$\frac{d\delta z}{d\tau} = -\eta \left( n_0 \delta z + z_0 \delta n + \delta n \delta z \right)$$
(A3)

For a linear stability analysis we write for the fluctuations $\delta n$ and $\delta z$:

$$\delta n(\tau) = \psi_n \exp(\lambda \tau)$$
$$\delta z(\tau) = \psi_z \exp(\lambda \tau)$$
(A4)

Eq.(A3) becomes up to the first order in the perturbations:

$$0 = (-\eta z_0 - \lambda)\psi_n - \eta n_0 \psi_z$$
$$0 = -\eta z_0 \psi_n + (-\eta n_0 - \lambda)\psi_z$$
(A5)

This linear set of relation has a non-trivial solution for:

$$0 = \lambda^2 + \lambda \eta \left( z_0 + n_0 \right)$$
(A6)

Hence the linear stability analysis delivers the eigenvalues:

$$\lambda = -\eta \left( z_0 + n_0 \right)$$
$$\lambda = 0$$
(A7)

While the negative eigenvalue indicates that the stationary state corresponds to a stable state the second zero valued eigenvalue suggests that the linear stability analysis is not sufficient to characterize the stationary state. However, if we take the higher order contribution of the form $-\eta\delta n\delta z$ in Eq.(A3) into account the model suggests that the stationary state given by Eq.(8) is also a stable state.

**Figures**

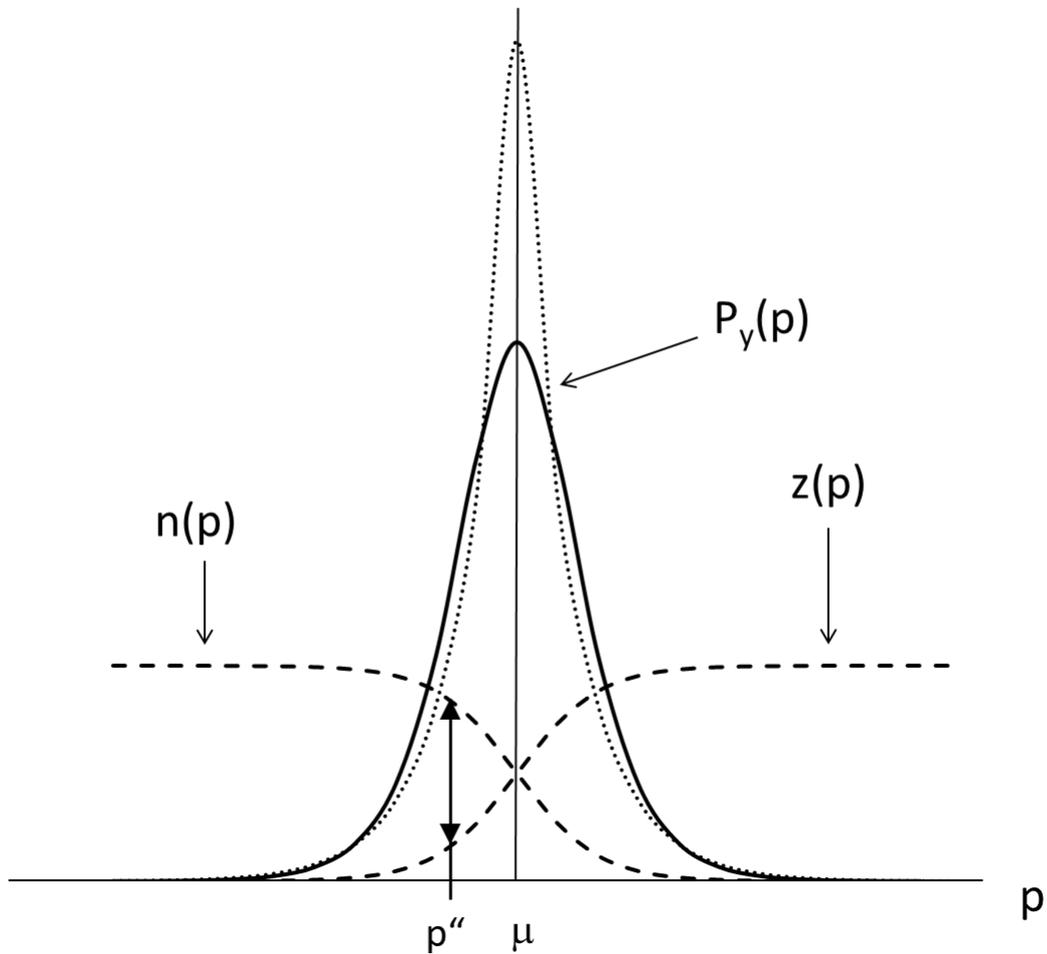

**Figure 1:** Schematically displayed is the price dependent number of desired units *n(p)* and the number of available units *z(p)* (dashed lines) together with the logistic price distribution $P_y(p)$ (fat line) which is symmetric around the mean price *μ*. The dotted line indicates the corresponding Laplace distribution.

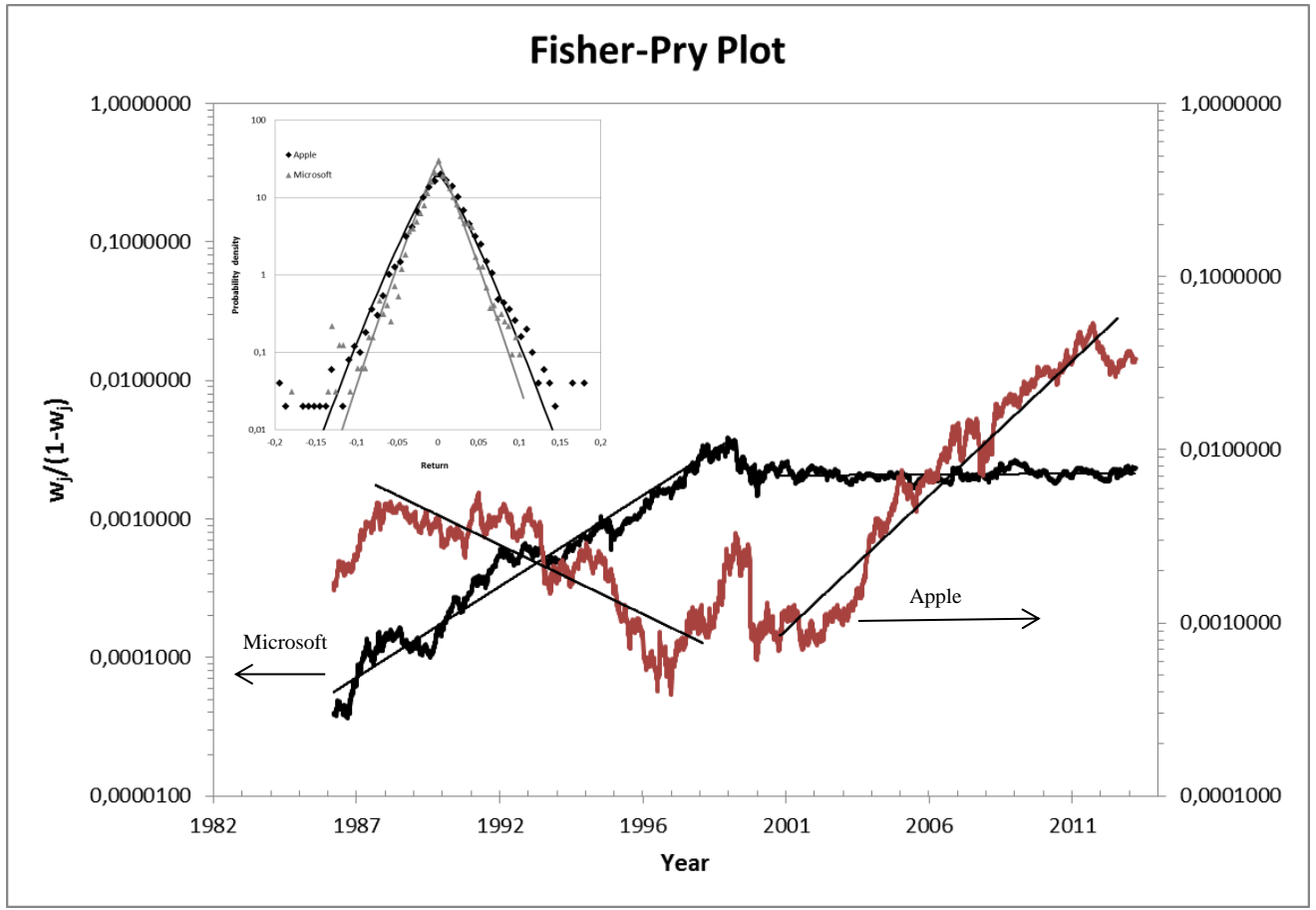

**Figure 2:** Fisher-Pry plot of the relative stock price $w_j$ of Microsoft and Apple stock shares in relation to the DJIA. The regression lines indicate competitive (dis-)advantages with respect to the DJIA. The insert shows the return distribution of both stocks while the central part is fitted by GED distributions.